%% file: main.tex
\documentclass[sigconf]{acmart}

\AtBeginDocument{%
  \providecommand\BibTeX{{%
    \normalfont B\kern-0.5em{\scshape i\kern-0.25em b}\kern-0.8em\TeX}}}

\setcopyright{acmcopyright}
\copyrightyear{2024}
\acmYear{2024}
\acmDOI{XXXXXXX.XXXXXXX}

\acmConference[Conference acronym 'XX]{30TH ACM SIGKDD CONFERENCE ON KNOWLEDGE DISCOVERY AND DATA MINING}{August 25--29,
  2024}{Barcelona, Spain}
\acmISBN{978-1-4503-XXXX-X/18/06}

\usepackage{amsfonts}
\usepackage{amsmath}
\usepackage{bbm}
\usepackage{color}
\usepackage{booktabs}
\usepackage{multirow}
\usepackage{xspace}
\usepackage{enumitem}
\usepackage{xcolor}
\usepackage{caption}
\usepackage{subcaption}
\usepackage[export]{adjustbox}
\usepackage[normalem]{ulem}
\usepackage[misc]{ifsym}
\useunder{\uline}{\ul}{}
\usepackage{makecell}

\def\modelName{LLM-KERec} 

\newcommand{\eg}{\emph{e.g.,}\xspace}

\newcommand{\ignore}[1]{}

\begin{document}

\title{Breaking the Barrier: Utilizing Large Language Models for Industrial Recommendation Systems through an Inferential Knowledge Graph}
\renewcommand{\shorttitle}{Utilizing Large Language Models for Industrial Recommendation Systems through an Inferential Knowledge Graph}

\author{Qian Zhao}
\affiliation{%
  \institution{Ant Group}
  \city{Shanghai}
  \country{China}}
\email{zq317110@antgroup.com}

\author{Hao Qian}
\affiliation{%
  \institution{Ant Group}
  \city{Shanghai}
  \country{China}}
\email{qianhao.qh@antgroup.com}

\author{Ziqi Liu}
\affiliation{%
  \institution{Ant Group}
  \city{Hangzhou}
  \country{China}}
\email{ziqiliu@antgroup.com}

\author{Gong-Duo Zhang}
\affiliation{%
  \institution{Ant Group}
  \city{Hangzhou}
  \country{China}}
\email{gongduo.zgd@antgroup.com}


\author{Lihong Gu}
\affiliation{%
  \institution{Ant Group}
  \city{Hangzhou}
  \country{China}}
\email{lihong.glh@antgroup.com}



\renewcommand{\shortauthors}{Qian Zhao and Hao Qian, et al.}

\begin{abstract}
  Recommendation systems are widely used in e-commerce websites and online platforms to address information overload. However, existing systems primarily rely on historical data and user feedback, making it difficult to capture user intent transitions. Recently, Knowledge Base (KB)-based models are proposed to incorporate expert knowledge, but it struggles to adapt to new items and the evolving e-commerce environment.
  To address these challenges, we propose a novel \textbf{L}arge \textbf{L}anguage \textbf{M}odel based Complementary \textbf{K}nowledge \textbf{E}nhanced \textbf{Rec}ommendation System ({\modelName}).
  It introduces an entity extractor that extracts unified concept terms from items and user information. To provide cost-effective and reliable prior knowledge, entity pairs are generated based on entity popularity and specific strategies. The large language model determines complementary relationships in each entity pair, and constructs a complementary knowledge graph. Furthermore, a new complementary recall module and an Entity-Entity-Item (E-E-I) weight decision model refine the scoring of the ranking model by using real complementary exposure-click samples.
  We conduct extensive experiments on three industry datasets. The results demonstrate significant performance improvement of our model compared to existing approaches. Additionally, detailed analysis shows that {\modelName} enhances users' enthusiasm for consumption by recommending complementary items.
  In summary, {\modelName} addresses the limitations of traditional recommendation systems by incorporating complementary knowledge and utilizing a large language model to capture user intent transitions, adapt to new items, and enhance recommendation efficiency in the evolving e-commerce landscape.

\end{abstract}




\received{20 February 2007}
\received[revised]{12 March 2009}
\received[accepted]{5 June 2009}

\maketitle

\input{introduction.tex}
\input{model.tex}

\input{experiment.tex}
\input{related_work.tex}
\input{conclusion.tex}
\newpage
\bibliographystyle{ACM-Reference-Format}
\bibliography{ref}

\appendix

\end{document}

%% file: introduction.tex
\section{Introduction}
The Recommendation System (RS) has been widely used in online service platforms (e.g., Amazon and Taobao) as an effective tool for alleviating information overload.
The primary objective of RS is to infer user preferences from their past behaviors, recommend the most suitable items that align with their interest.
Hence, existing recommendation systems are mostly trained based on historical exposure and click logs.

Hereby, we summarize the existing recommendation tasks as the combination of following sub-tasks:
1) Recommend substitutive items based on the exposure or click feedback from users. 
2) Recommend complementary items based on the conversion feedback from users. 
3) Conduct traffic exploration or business intervention to explore users' other potential interests.
Traditional deep Click-Through Rate (CTR) prediction models\cite{wang2017deep,lian2018xdeepfm,huang2019fibinet,wang2021masknet} equipped with well-designed feature interaction techniques through deep neural networks
have been widely applied to tackle these sub-tasks in major e-commerce systems. 
These methods provide personality in RS via extracting user preference from historical exposure-click samples. 
Despite achieving notable performance improvements in RS, we argue they still suffer from the following two major challenges in real-world scenarios.
1) These models rely heavily on exposed samples and user feedback, which limits the performance of RS in cold-start scenarios and makes it difficult to cope with the continuous emergence of new items.
2) The sparsity of user interaction samples results in existing CTR models being more effective in recommending substitutes (sub-task 1) than complementary items (sub-task 2).
While models based on expert-crafted complementary rules or knowledge graphs can aid in recommending complementary items, they are not a sustainable solution in the ever-evolving landscape of e-commerce due to efficiency and expenditure challenges. 
Therefore, it's indispensable to incorporate efficient knowledge and Large Language Model (LLM) as the carrier of human reasoning and logic to improve the performance of RS\cite{devlin2018bert,10.5555/3495724.3495883,du2022glm,touvron2023llama}.
However, in RS, due to the difficulty of large-scale deployment and long inference time of the large language model,
it has only been used as a tool for text embedding in previous work\cite{chen2017joint,hui2022personalized}, making it difficult to fully utilize its powerful reasoning ability.

In the light of the above limitations and challenges, we propose a novel {\modelName} for recommendation.
Our method combines the efficient collaborative signal processing capability of traditional models with large language models and complementary graph to help users quickly find their preferred items.
This method not only reduces the homogeneity of traditional model recommendation results, but also improves overall click-through and conversion rates.
Specifically, we first use our designed entity extractor to extract unified concept terms (referred to as entities) from the information of all items and user billing information.
Next, we generate entity pairs based on the popularity of entities and carefully designed strategies.
Then we construct a complementary graph based on a large language model, where each edge in the graph represents a complementary purchasing relationship between corresponding entities.
Finally, we launch a new complementary recall module and train the E-E-I weight decision model through real exposure click samples.
This model will apply the edge weights of the graph corrected by real feedback to the fine-ranking layer model to achieve recommendation of complementary items.
It is worth mentioning that both the entity extractor and complementary graph are periodically updated to adapt to new items and the changing e-commerce environment.
The main contributions of this paper can be summarized as follows:
\begin{itemize}
    \item For the first time, we utilize the inference ability of large language models as a medium to improve the scenario preference when recommending items to each user, achieving large-scale application of large language model in industrial scenarios.
    \item Our method continuously adjusts the weights of graph edges based on real exposure samples of complementary item pairs, addressing the language model's weakness in determining user preference strength.
    \item Extensive experiments are conducted on three industry scenarios, demonstrating our approach is consistently better than a number of competitive baselines.
\end{itemize}

%% file: model.tex
\section{System Overview}
\label{section:methodology}
\begin{figure}
    \centering
    \includegraphics[width=0.48\textwidth]{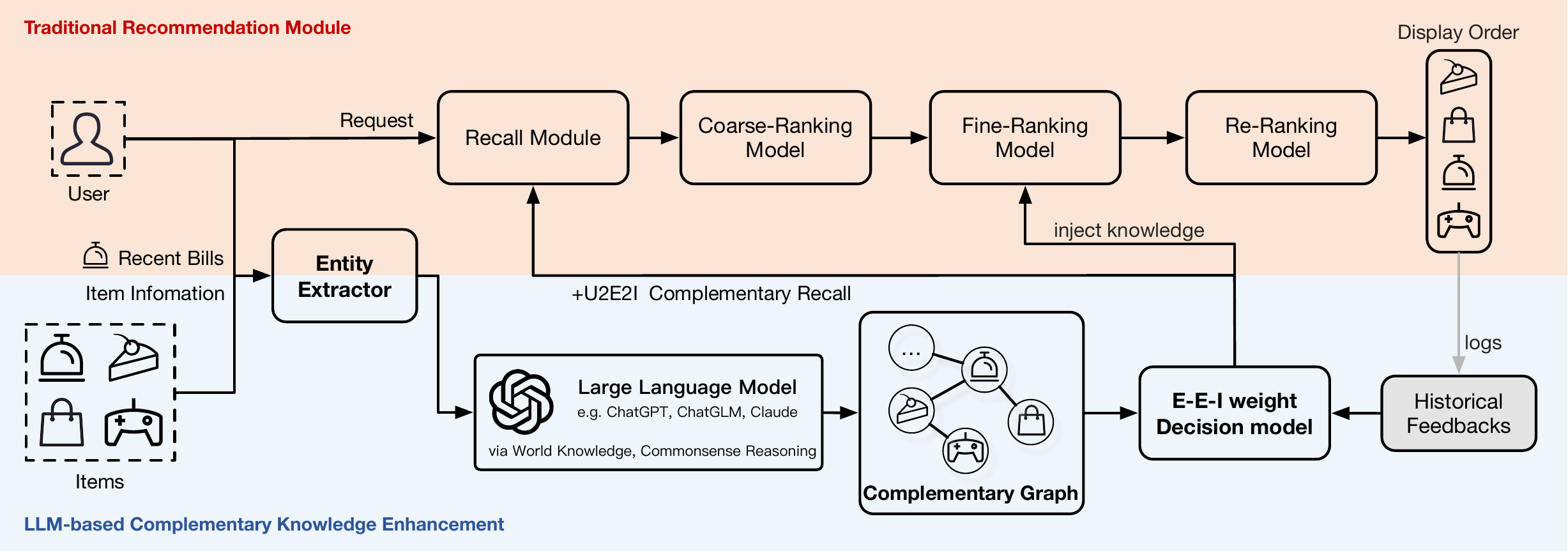}
    \caption{Overall framework of our proposed {\modelName} System.}
    \label{fig:model_framework}
\end{figure}

In this section, we present the overview of the {\modelName} System, including \emph{Traditional Recommendation Module} and \emph{LLM-based Complementary Knowledge Enhancement}, 
shown in Fig. \ref{fig:model_framework}.

\subsection{Traditional Recommendation Module}

In the traditional recommendation architecture, when a user opens an application, the application will automatically send a request to server. This process follows these steps:
1) The server triggers the recall module based on the user's request information, including popular item recall, LBS recall, personalized recall, etc. The recall module returns a large number of candidate items.
2) These candidate items are then input into the coarse-ranking model for filtering. The coarse-ranking model produces a smaller set of candidate items.
3) Finally, the fine-ranking model and re-ranking model are used to make the final decision on the display order of these items.
Additionally, manual intervention may occur at each step, such as assigning weights to the items for publication.

The fine-ranking model and re-ranking model are typically trained using historical exposure and click logs. As a result, existing recommendation models often prioritize recommending similar items based on positive user feedback. This poses a challenge when it comes to providing reliable recommendations for supplementary items that have potential reasoning behind them, such as suggesting complementary item B after a user has purchased item A.




\subsection{LLM-based Complementary Knowledge Enhancement}
In this paper, the {\modelName} system maintains the ability to efficiently process a large number of collaborative signals of the existing recommendation system. It also overcomes above challenge through the LLM-based Complementary Knowledge Enhancement Module.
To establish connections between different content in Alipay, {\modelName} creates a unified entity (category) system for users' billing behaviors and all items. Each item or bill is classified into a unique entity, which serves as a bridge between various contents. Utilizing world knowledge and commonsense knowledge, we employ a large language model to determine the existence of a complementary relationship between two entities and construct a complementary graph.
The nodes of this graph are all entities, while the edges indicate the complementary relationship between the corresponding entities. Subsequently, using the real exposure and click feedback of complementary items, we train an entity-entity-item (E-E-I) weight decision model. This model is then used to inject knowledge into the ranking model.
By adopting this approach, we can provide personalized recommendations for both favorite items and complementary items. This solution has been successfully implemented in Alipay marketing scenarios, and experimental results have demonstrated its effectiveness.



\section{Diving into the {\modelName} System}
In this section, we will zoom into each module in {\modelName} System.
\subsection{Entity Extractor}
\label{section:entity_extractor}
\subsubsection{Entity Dict}

In real-world applications, like Alipay, users' behaviors span across various scenarios, each with diverse content. To align information and knowledge from these diverse sources, it is crucial to establish a unified association pattern. This is where our Entity Dict comes into play, serving as a bridge for different content types.
In the Entity Dict, each entity represents a specific concept, such as ``phone'' or ``cola''. Our dedicated group of experts meticulously designed the Entity Dict, incorporating tens of thousands of entities. Importantly, the Entity Dict is regularly updated every week to ensure its adaptability to new items and content.

\subsubsection{Extracting Entities}

\begin{figure}
    \centering
    \includegraphics[width=0.47\textwidth]{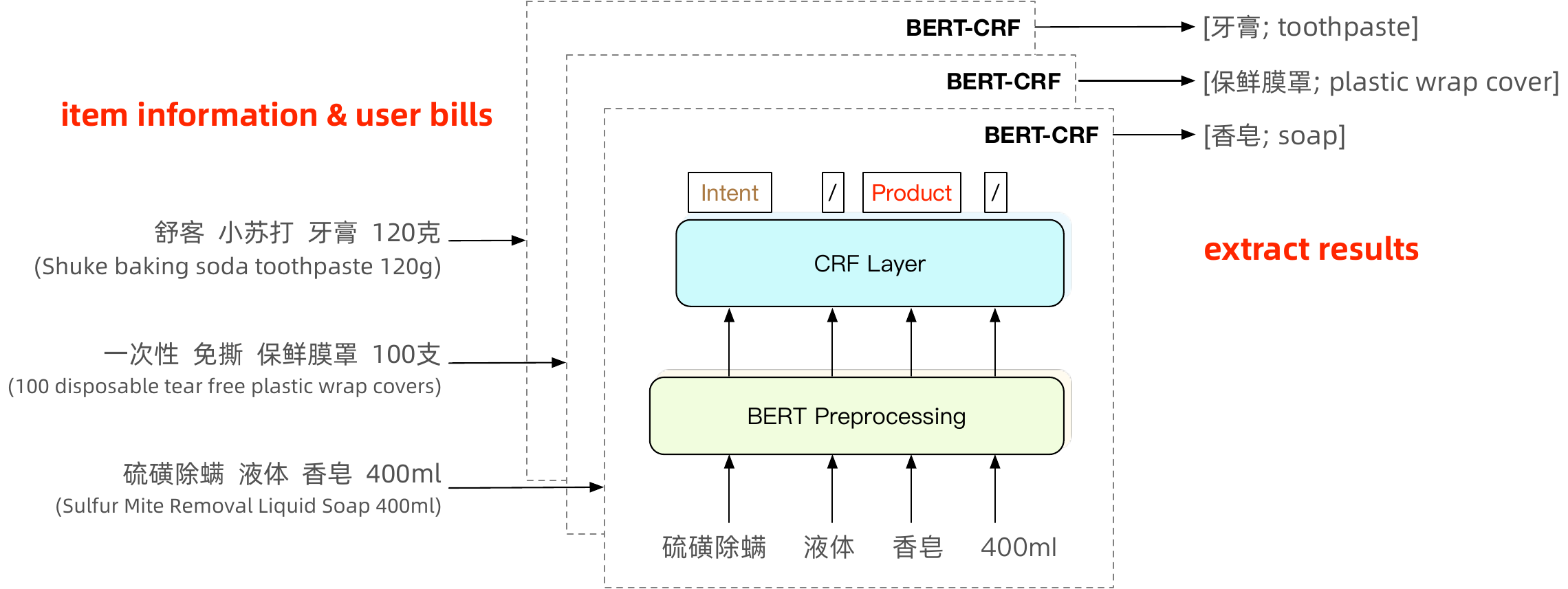}
    \caption{Extracting entites from item information and user bills.}
    \label{fig:extracting_entities_bert_crf}
\end{figure}

Building upon the Entity Dict, our focus shifts to extracting entities from various user behaviors within Alipay, including bills, visit logs, and the entity information of items in marketing scenarios. This extraction process can be viewed as a Named Entity Recognition (NER) task, which has been extensively studied in the field of Natural Language Processing (NLP) \cite{zhang-yang-2018-chinese,liu-etal-2019-encoding,10.1109/TKDE.2020.2981314}.
To perform entity extraction, we utilize the BERT-CRF model. This model combines the transfer capabilities of BERT\cite{devlin2018bert} with the structured predictions of CRF\cite{10.5555/645530.655813}. The BERT-CRF model enables us to accurately extract entities from user behaviors in Alipay.
In the LLM-based Complementary Knowledge Enhancement, our primary objective is to establish connections between user purchase behaviors and the items to be recommended. To achieve this, we extract entities from each user's recent bills, forming their recent entity transaction sequence. Furthermore, we extract entities from item information and assign a unique entity as the item's category. The detailed procedure is illustrated in Fig. \ref{fig:extracting_entities_bert_crf}.

\subsection{Complementary Graph Construction}

We utilize the results of the entity extractor to construct a complementary graph, which helps us gain insights into users' purchasing patterns. Specifically, we aim to understand which item \emph{B} (\eg paper towels) users typically buy after purchasing item \emph{A} (\eg utensils) by leveraging natural language understanding and commonsense reasoning.
The construction of the complementary graph involves two main steps:
1) We generate candidate entity pairs from the entity dict, ensuring both execution efficiency and comprehensive item coverage.
2) Through a combination of carefully designed prompt engineering and the utilization of a large language model, we perform reasoning tasks to extract meaningful insights from the data.

\subsubsection{Entity Pair Construction}

\begin{figure}
    \centering
    \includegraphics[width=0.475\textwidth]{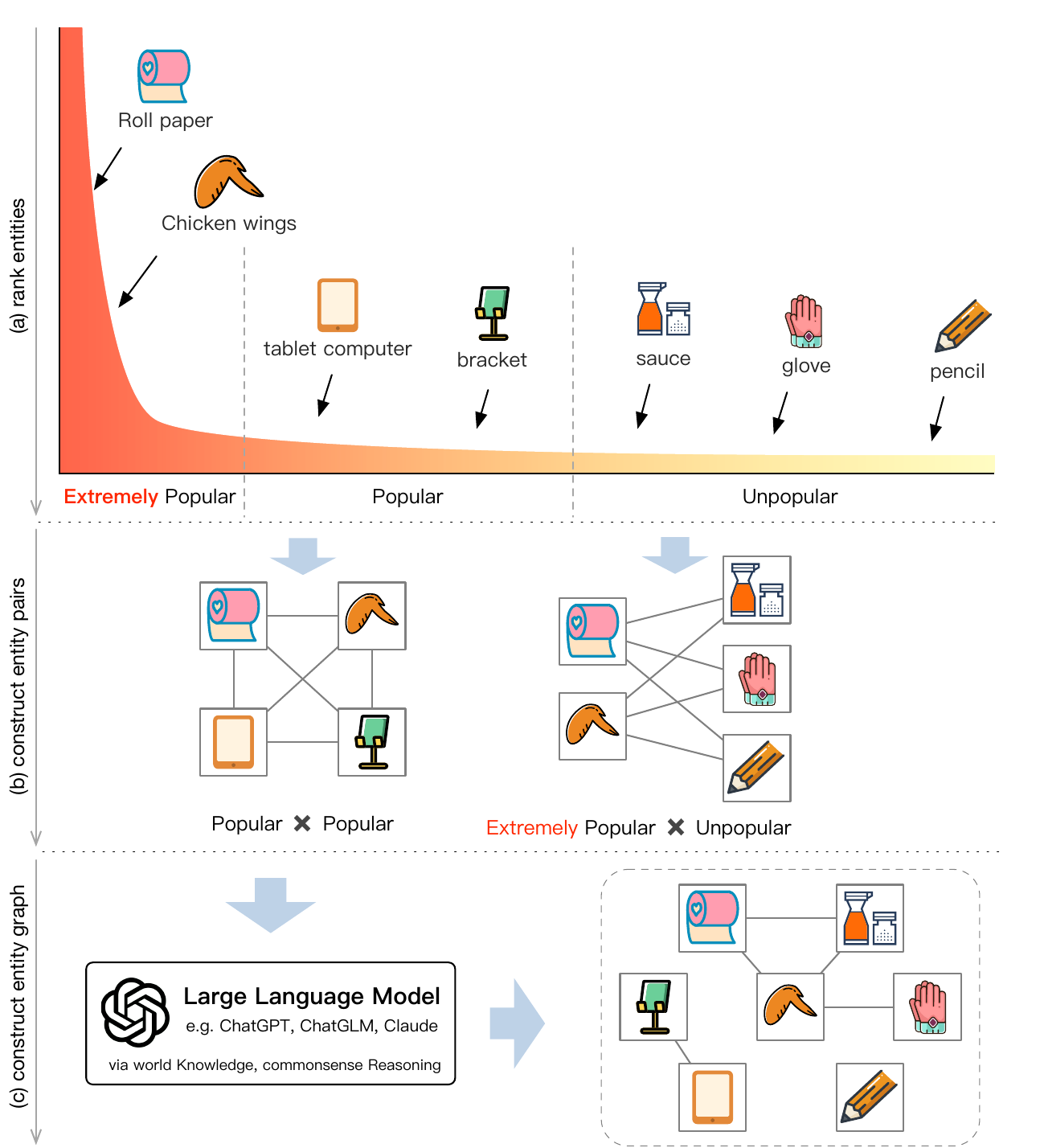}
    \caption{Long-tail distribution in Entity Dict.}
    \label{fig:long-tail-dist}
\end{figure}

Firstly, it is important to recognize that certain items have complementary relationships with specific concepts, and these concepts often encompass more specific items. In industrial e-commerce scenarios, where the number of items can reach millions or even more, there are only a few thousand concept categories. By using concepts as entities instead of individual items, computational resources can be significantly conserved. In Section \ref{section:entity_extractor}, we have already assigned unique entities to all items using the entity extractor.
To construct entity pairs, a straightforward approach would involve taking elements from a set containing $n$ entities and combining them pairwise, resulting in $\frac{n(n-1)}{2}$ candidate entity pairs. However, this method is not cost-effective due to the slower inference speed of downstream large language models. Additionally, in real-world scenarios, there is often a long-tail distribution where a few entities are frequently purchased while the majority of entities are rarely consumed (as depicted in Fig. \ref{fig:long-tail-dist}). Focusing solely on tail entity combinations makes it challenging to improve the overall performance of the recommendation system.


To tackle this challenge, we have devised a cost-effective segmented combination strategy as follows:
1) Initially, we sort entities in descending order based on metrics like total conversions and clicks. This allows us to classify them into three categories: extremely popular, popular, and unpopular entities.
2) We focus on constructing entity pairs exclusively within the popular entities. This approach enhances the performance and coverage specifically for popular items.
3) Additionally, we construct entity pairs that include both extremely popular and unpopular entities. This ensures comprehensive coverage in the complementary graph for unpopular items.
By merging and eliminating duplicates from all entity pairs, we obtain the final output.
This segmented combination strategy ensures reliable support for downstream modules while minimizing resource waste.

\subsubsection{Large Language Model}

Large language models have garnered significant attention from researchers due to their remarkable understanding and reasoning abilities in natural language processing. A specific research direction, exemplified by methods like Prompt-Turing\cite{lester-etal-2021-power} and LoRA\cite{hu2022lora}, explores prompt engineering techniques based on these large language models.
In these approaches, researchers can obtain desired answers from the large language model by providing a simple task description and a small number of examples. By fine-tuning the model efficiently using techniques like LoRA\cite{hu2022lora} based on annotated samples, researchers can enhance its support for the current task.
In this study, we also leverage the capabilities of large language models to determine the existence of a complementary relationship in an entity pair. Specifically, we utilize Claude 2\footnote{https://www.anthropic.com/index/claude-2} as the underlying language model and thoughtfully design reliable prompts to guide the model in conducting a step-by-step analysis and providing dependable reasoning evidence. The ultimate goal is to enhance the interpretability of the reasoning results.
Upon completing the reasoning process, we sample thousands of examples for manual annotation and continuously refine the prompts to attain an acceptable level of accuracy in the reasoning outcomes.


The prompts we have designed encompass various aspects, including:
1) Description of the input data format, where each line consists of two entities representing real-world concepts.
2) Task description, which involves determining whether there is a likelihood of a person purchasing entity \emph{B} shortly after purchasing entity \emph{A}.
3) Provide multiple data examples and their corresponding reasons. For instance, we provide examples like the complementary relationship between bread and milk, as they form a popular breakfast combination. Conversely, we highlight that there is no complementary relationship between a phone and milk, as they are unrelated.
4) Explanation of the output format, which includes a concise description of the purposes of the two entities, whether a complementary relationship exists between them, and a detailed explanation. Ultimately, the answer is denoted as either Y or N.
Moreover, we have explored methods such as ChatGPT 3.5\footnote{https://openai.com/blog/chatgpt} and ChatGLM 2\cite{du2022glm,zeng2022glm}. A comprehensive comparison between these methods can be found in Section \ref{llm_exp}.

\subsubsection{Automatic Update Strategy}

In a real e-commerce environment, users and merchants continually rely on each other's cognitive updates and mutually promote one another. This means that the popularity of entities is not static. For instance, certain merchants may employ marketing strategies to rapidly gain public attention for their products, and over time, older products may be phased out.
To address this dynamic nature of popularity, we have implemented an automatic daily schedule for constructing the incremental complementary graph. By promptly recognizing such changes and updating our complementary graph accordingly, we can ensure the effective and sustained operation of the entire system. This proactive approach is crucial for maintaining optimal system performance in the long run.

\subsection{E-E-I weight decision model}

At present, we have successfully linked each user's recent bills and each item to entities in the complementary graph. 
Our objective is to recommend complementary items (entity2) based on user bill (entity1), where relation \emph{entity1}-\emph{entity2} exist in the complementary graph.
However, due to the limited ability of LLM to accurately assess user preferences, we require an E-E-I (\emph{entity1-entity2-item}) weight decision model to effectively accomplish this task.

\subsubsection{Model Overview}

Intuitively, the success of the {\modelName} System relies heavily on the construction of a high-quality E-E-I weight decision model. Therefore, we propose a Two-stage Complementary Knowledge Enhancement Procedure, which consists of the \emph{Ranking Stage} and the \emph{Integration Stage}, as shown in Fig. \ref{fig:pipeline}.
In the following sections, we will take a closer look at each well-designed stage.

\subsubsection{Ranking Stage}

As shown in Fig. \ref{fig:pipeline}(a.0), our model adopts a dual-tower architecture, where the outputs of the two towers represent the representations of the complementary item and bill entity, respectively. The dot product of these outputs serves as the preference level indicator. For the representation of item, we can extract a rich set of features from the database, including basic features, statistical features, and interaction features, etc.
However, for the entity representation, we face a challenge as we lack specific information to describe them, aside from a pre-assigned ID. To overcome this limitation, we employ Graph Neural Network\cite{kipf_semi-supervised_2019} and Contrastive Learning to  representative entity from two distinct perspectives: the first-order substitutable view and the second-order complementary view.
The Ranking Stage can be further subdivided into the following modules:
\begin{figure*}
    \centering
    \includegraphics[width=0.95\textwidth]{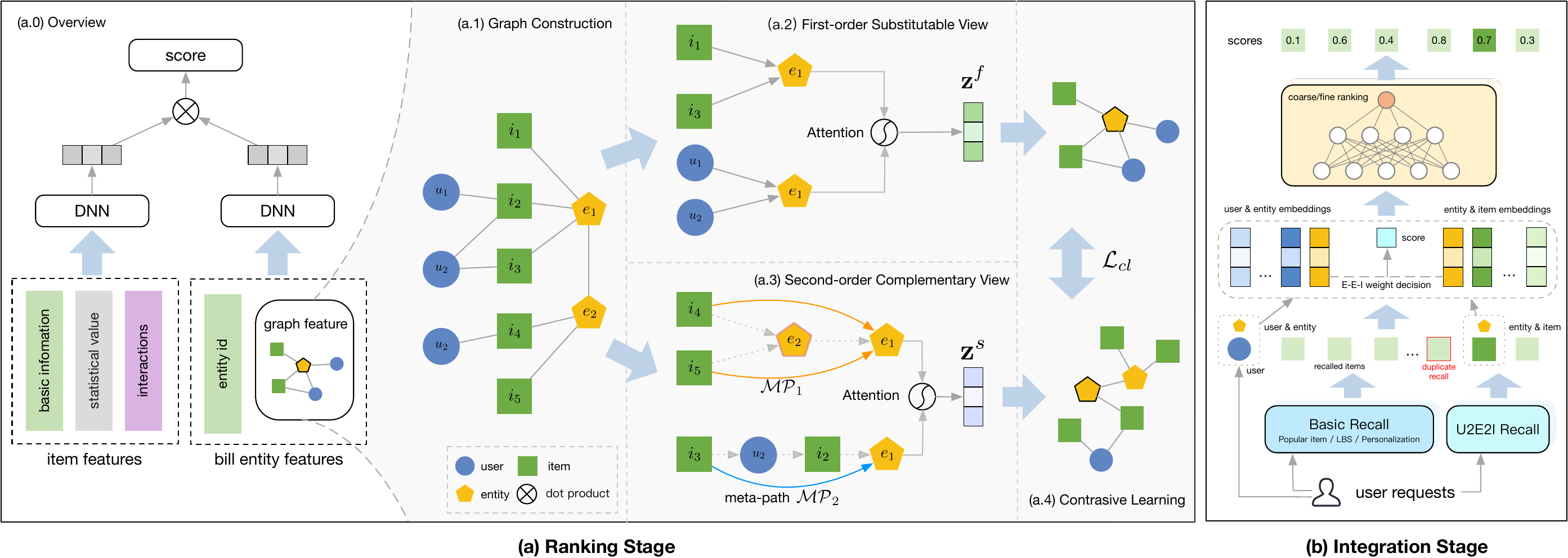}
    \caption{Overall framework of Two-stage Complementary Knowledge Enhancement Procedure.}
    \label{fig:pipeline}

\end{figure*}

\paragraph{Graph Construction}
Graph Neural Networks (GNNs) has demonstrated promising results for recommender systems, as they can effectively leverage high-order relationship.
These methods represent interaction data as graphs, such as the user-item interaction graph, and iteratively propagate neighborhood information to learn effective node representations.
Similarly, as show in Fig. \ref{fig:pipeline}(a.1), we have designed the following edge relationships around how to better represent entity:
1) Establish edges for click behaviors between user nodes and item nodes.
2) Establish edges for dependency relationships between item nodes and entity nodes.
3) Establish edges for complementary relationships between entity nodes and entity nodes.

Given the user set $\mathcal{U} = \{u\}$, the item set $\mathcal{I} = \{i\}$ and the entity set $\mathcal{E} = \{e\}$.
The number of nodes is $n = |\mathcal{U}| + |\mathcal{I}| + |\mathcal{E}|$.
Our method formulate the available data as a user-item-entity graph $\mathcal{G} = (\mathcal{V}, \mathbf{A})$,
where $\mathcal{V} = \mathcal{U} \cup \mathcal{I} \cup \mathcal{E}$ and $\mathbf{A} \in \mathbb{R}^{n \times n}$ is the adjacent matrix.

\paragraph{First-order Substitutable View}
\label{section:first-order-substitutable-view}
In order to model substitutable relationships, we consider two different sources of information for each entity:
\begin{enumerate}
    \item From an item sub-perspective, we need to explore the common features of items that have a dependency relationship on the current entity.
    \item Similarly, from an user sub-perspective, we need to explore the common features of the users who frequently click on the current entity.
\end{enumerate}
Specifically, we aggregate information using the Graph Attention Network (GAT), denoted by $\mathbf{h}_i' = f_t(\mathbf{h}, i, \mathcal{N}_i; \theta_t)$.
Here, $\mathbf{h}$ represents the embeddings of all nodes, $i$ denotes current node index, $\mathcal{N}_i$ is neighbors of node $i$, $\theta_t$ is the network parameters,
and function $f(\cdot, \cdot, \cdot; \cdot)$ is defined as:
\begin{equation}
    \label{eq:gat_level1}
    \mathbf{h}_i' = \sum_{j \in \mathcal{N}_i} \sigma\Big( \alpha_{ij} \mathbf{W}_1 \mathbf{h}_{j} \Big),
\end{equation}
where $\alpha_{ij}$ is defined as:
\begin{equation}
    \label{eq:gat_level2}
    \alpha_{ij} = \frac{\exp\Bigg(\textsc{LeakyReLU}\Big(\mathbf{W}_2[\mathbf{W}_1\mathbf{h}_i || \mathbf{W}_1\mathbf{h}_j] \Big)\Bigg)}{\sum_{k \in \mathcal{N}_i} \exp\Bigg(\textsc{LeakyReLU}\Big(\mathbf{W}_2[\mathbf{W}_1\mathbf{h}_i || \mathbf{W}_1\mathbf{h}_k]\Big)\Bigg)},
\end{equation}
where $\mathbf{h}_i$ represents the embeddings of node $i$,
$\mathbf{W}_1 \in \mathbb{R}^{d \times d}$ and $\mathbf{W}_2 \in \mathbb{R}^{2d}$ are trainable parameters,
$\sigma(\cdot)$ is a non-linearity activate function, $\textsc{LeakyReLU}(\cdot)$ is the LeakyReLU activate function
and $[\cdot || \cdot]$ is concatenate operation.
Then we can fuse information from different sub-perspectives including user and item side, based on attention mechanism to obtain entity node $i$ embedding:
\begin{equation}
    \label{eq:gat_level3}
    \mathbf{z}_i^{f} = \textsc{Attention}\Big(f_t(\mathbf{h}, i, \mathcal{N}_i \cap \mathcal{I}; \theta_1), f_t(\mathbf{h}, i, \mathcal{N}_i \cap \mathcal{U}; \theta_1)\Big).
\end{equation}
As shown in the Fig. \ref{fig:pipeline}(a.2), where $i_1$ and $i_3$ are aggregated to $e_1$ on the item side, and $u_1$ and $u_2$ are aggregated to $e_1$ on the user side.

\paragraph{Second-order Complementary View}
In the modeling of complementary relationships, we also consider two different sources of information for each entity:
\begin{enumerate}
    \item From the complementary graph, we design a meta path $\mathcal{MP}_1$: \emph{item (database) -> entity (graph) -> entity (bill)}, represents the collection of item features complementary to the current entity from the perspective of semantic reasoning.
    \item From the user' daily behaviors, we also design a meta path $\mathcal{MP}_2$: \emph{item1 (bill) -> user -> item2 (bill) -> entity (bill)}, which indicates what items have been recently consumed by users who have consumed \emph{item2} in the short term.
\end{enumerate}
Similarly, we obtain the representation of entity node $i$ through Eq. \ref{eq:gat_level4}:
\begin{equation}
    \label{eq:gat_level4}
    \mathbf{z}_i^{s} = \textsc{Attention}\Big(f_t(\mathbf{h}, i, \mathcal{N}_{i-\mathcal{MP}_1}; \theta_2), f_t(\mathbf{h}, i, \mathcal{N}_{i-\mathcal{MP}_2}; \theta_2)\Big),
\end{equation}
where $\mathcal{N}_{i-\mathcal{MP}_1}$ and $\mathcal{N}_{i-\mathcal{MP}_2}$ are the target node sets explored throught meta-path $\mathcal{MP}_1$ and $\mathcal{MP}_2$, respectively, starting from entity node $i$.
As shown in the Fig. \ref{fig:pipeline}(a.3), where $i_4$ and $i_5$ are aggregated to $e_1$, and $i_3$ is aggregated to $e_1$.

\paragraph{Contrasive Learning}
$\mathbf{z}_i^f$ and $\mathbf{z}_i^s$ are aggregated through information from \emph{first-order substitutable view} and \emph{second-order complementary view}, respectively,
representing the characterization of entity $i$ from two independent and complementary perspectives.
$\mathbf{z}_i^f$ and $\mathbf{z}_i^s$ are interrelated and complementary, as they can supervise each other in training process.
Therefore, we utilize the contrastive loss, InfoNCE~\cite{oord2018representation}, to maximize the agreement of positive pairs and minimize that of negative pairs:
\begin{equation}
    \label{eq:info_nce}
    \mathcal{L}_{cl} = \sum_{i\in\mathcal{E}} -\log \frac{\exp(s(\mathbf{z}_i^f, \mathbf{z}_i^s) / \tau)}{\sum_{j\in\mathcal{E}} \exp(s(\mathbf{z}_i^f, \mathbf{z}_j^s) / \tau)},
\end{equation}
where $s(\cdot)$ measures the similarity between two vectors, which is set as cosine similarity function; $\tau$ is the hyper-parameter, known as the temperature in softmax.
Finally, the representation of the node $i$ is the weighted sum of $\textbf{z}_i ^ {f} $and $\textbf{z}_i ^ {s} $, which will be used for downstream recommended tasks.

\paragraph{Training Process}
We leverage a multi-task training strategy to optimize the main E-E-I weight decision task and the auxiliary tasks including contrastive learning task and L2 normalization task jointly:
\begin{equation}
    \mathcal{L} = \mathcal{L}_{main} + \lambda_1 \mathcal{L}_{cl} + \lambda_2 ||\Theta||^2,
\end{equation}
where $\Theta$ is the set of model parameters, $\lambda_1, \lambda_2$ are hyperparameters to control the strengths of the diversity preserving loss.
$\mathcal{L}_{main}$ is the Cross Entropy Loss of the main E-E-I weight decision task.

\subsubsection{Integration Stage}

To effectively and efficiently recommend items complemented by recent user bills to those with higher demand,
we optimize both the recall module and the fine-ranking model, as shown in Fig. \ref{fig:pipeline}(b).
Specifically, for the recall module, we added a new complementary recall route.
To avoid excessive recall, we prepared a set of up to top-k newly recalled complementary items based on the scores from the \emph{E-E-I weight decision model} and the recent bill entities retrieved from real-time requests by user ID.
As for the fine-ranking model, during the training phase, we also introduce the \emph{E-E-I weight decision model} to provide scores, entity embeddings, and item embeddings for current samples.
The new recall module enables the downstream fine-ranking model to pay attention to complementary items, 
overcoming the limited input of complementary items caused by exposure bias in previous recommendation systems.
The fine-ranking model combines the features of current complementary items and user profile behaviors to comprehensively and personalized sort candidate items.

%% file: experiment.tex



\section{Experiments}
\label{section:experiments}
To verify the effectiveness of the proposed {\modelName}, we conduct extensive offline experiments utilizing the real industrial dataset procured from the Alipay online environment and report detailed analysis results. Moreover, we conduct online A/B tests in real-world marketing recommendation scenario to evaluate the performance of {\modelName} in real industrial applications.

This section encompasses a series of experiments designed to answer the following key questions:
\begin{itemize}

  \item \textbf{Q1}: How does {\modelName} perform when compared with other state-of-the-art (SOTA) baseline methods? (see Subsection \ref{off_comparison_exp})
  \item \textbf{Q2}: How does {\modelName} perform in real-world industrial applications? (see Subsection \ref{on_comparison_exp})
  \item \textbf{Q3}: How do the distinct modules of {\modelName} contribute to performance improvements? (see Subsection \ref{ablation_exp})
  \item \textbf{Q4}: How do the different large language models impact the performance of {\modelName}? (see Subsection \ref{llm_exp})
\end{itemize}

\subsection{Experimental Setups}
\subsubsection{Datasets}
This paper mainly focuses on recommendation in digital marketing scenarios, we utilize real-world industrial datasets\footnote{The data set does not contain any Personal Identifiable Information (PII). The data set is desensitized and encrypted. Adequate data protection was carried out during the experiment to prevent the risk of data copy leakage, and the data set was destroyed after the experiment.} from Alipay. It includes three major marketing and recommendation scenarios within Alipay: Super 567 (Dataset A), Consumer Channel (Dataset B), and Payment Result Page (Dataset C).
The Alipay application (APP) facilitates the presentation of numerous coupons to users through Super 567 and the Payment Result Page. The intention is to encourage user engagement by prompting them to click and collect these coupons, subsequently redeeming them through purchases made on Alipay. Moreover, within the Consumer Channel, the APP directly showcases goods that align with users' potential interests, aiming to stimulate clicks and subsequent purchases. Each day, a substantial user base, amounting to tens of millions, is exposed to the assortment of coupons and goods available on Alipay. To conduct our study, we randomly selected some instances spanning various dates over a one-month duration. The primary objective underlying data optimization efforts is to augment user conversions.

These scenarios exhibit significant differences in terms of user population distribution, as well as user intentions and behaviors. They are further randomly divided into the disjoint training set, validation set, and test set. The statistics of these datasets are presented in Table \ref{tab:dataset_info}.

\begin{table}[]
  \centering
  \setlength\tabcolsep{5pt} 
  \caption{The statistics of three datasets.}
  \label{tab:dataset_info}
  \begin{tabular}{@{}lllllll@{}}
    \toprule
              & \#Users  & \#Items & \#click  & \#conversion \\ \midrule
    Dataset A & 155209   & 84846   & 285001   & 26780        \\
    Dataset B & 1301782  & 376111  & 1448213  & 19437        \\
    Dataset C & 28361313 & 172786  & 15336011 & 143502       \\ \bottomrule
  \end{tabular}

\end{table}


\subsubsection{Evaluation Metrics}
In order to assess the overall system performance, we employ AUC (Area Under Curve) as the evaluation metric for offline experiments. Despite the actual industrial scenario being a ranking scenario, we simplify the offline experiments by treating them as a binary classification problem during modeling. In this approach, the model produces a score indicating whether the user likes (clicks or converts) the recommended item. Therefore, AUC is utilized for offline evaluation purposes. For online experiments, we directly measure the quality of different models by counting the number of clicks and conversions made by real users in various experimental groups. Consequently, the experimental group exhibiting a higher number of recommended items clicked and converted by users signifies better model performance.

\subsubsection{Baselines}
We choose the state-of-the-art recommendation system models as baselines for efficiency comparison. The baselines include
\textbf{DNN}\cite{10.5555/521706},
\textbf{Wide\&Deep}\cite{cheng2016wide},
\textbf{DCN}\cite{wang2017deep},
\textbf{ESMM}\cite{ma2018entire},
\textbf{PLE}\cite{DBLP:conf/recsys/TangLZG20} and \textbf{Masknet}\cite{wang2021masknet}.

\subsection{Offline Performance Comparison} \label{off_comparison_exp}
Table \ref{tab:experiment_result_all} presents the AUC results of the offline performance comparison for all methods. The Click and Conv. columns indicate the click AUC and conversion AUC values for the three datasets, respectively. In order to facilitate a more comprehensive comparison, we have incorporated the i-i graph into the baseline models, denoted as "+ ii graph". This adjustment has been implemented as our methodology capitalizes on graph-based techniques. The superior results are emphasized in bold, while the second-best results are denoted in underline. We utilize the symbol $``\dagger"$ to indicate that {\modelName} exhibits a significant difference from the top-performing baseline, as determined by paired t-tests at a significance level of $0.01$. Upon meticulous examination of the table, it is evident that {\modelName} surpasses other methods in terms of AUC across the three datasets, exhibiting superior performance across the majority of experimental outcomes.

\begin{table}[]
  \centering
  \caption{Offline performance comparison, with evaluation metrics including click AUC and conversion AUC(conv.). The best results are bolded and the second best results are underlined.}
  \label{tab:experiment_result_all}
  \resizebox{0.48\textwidth}{!}{
    \begin{tabular}{@{}lllllll@{}}
      \toprule
                          & \multicolumn{2}{c}{Dataset A} & \multicolumn{2}{c}{Dataset B} & \multicolumn{2}{c}{Dataset C}                                                                                        \\
                          & Click                         & Conv.                         & Click                         & Conv.                      & Click                      & Conv.                      \\ \midrule\midrule
      DNN                 & 0.61182                       & 0.75844                       & 0.77597                       & {\ul 0.76092}              & {\ul 0.86060}              & 0.93010                    \\
      DNN + ii graph      & {\ul 0.61580}                 & {\ul 0.80684}                 & {\ul 0.77751}                 & 0.73187                    & \textbf{0.86061}           & {\ul 0.93997}              \\
      \textbf{\modelName} & \textbf{0.62882}$^\dagger$    & \textbf{0.82460}$^\dagger$    & \textbf{0.78523}$^\dagger$    & \textbf{0.76271}$^\dagger$ & 0.85972                    & \textbf{0.94608}           \\ \midrule
      Wnd                 & 0.60599                       & 0.72751                       & {\ul 0.77766}                 & 0.74243                    & 0.86025                    & 0.93069                    \\
      Wnd + ii graph      & {\ul 0.62822}                 & {\ul 0.81782}                 & 0.77507                       & {\ul 0.75197}              & \textbf{0.86064}           & {\ul 0.93384}              \\
      \textbf{\modelName} & \textbf{0.63207}$^\dagger$    & \textbf{0.81896}$^\dagger$    & \textbf{0.77897}$^\dagger$    & \textbf{0.77140}$^\dagger$ & {\ul 0.86059}              & \textbf{0.94064}$^\dagger$ \\ \midrule
      DCN                 & 0.62457                       & {\ul 0.81760}                 & {\ul 0.78053}                 & 0.75231                    & 0.84966                    & 0.92919                    \\
      DCN + ii graph      & {\ul 0.63121}                 & 0.80487                       & 0.778024                      & {\ul 0.75280}              & {\ul 0.85706}              & {\ul 0.93370}              \\
      \textbf{\modelName} & \textbf{0.67284}              & \textbf{0.82507}$^\dagger$    & \textbf{0.78285}              & \textbf{0.76789}$^\dagger$ & \textbf{0.85732}$^\dagger$ & \textbf{0.94174}$^\dagger$ \\\midrule
      ESMM                & 0.61259                       & 0.78366                       & 0.76509                       & 0.75246                    & 0.85090                    & 0.91357                    \\
      ESMM + ii graph     & {\ul 0.61927}                 & {\ul 0.82100}                 & {\ul 0.77920}                 & {\ul 0.76136}              & {\ul 0.85378}              & {\ul 0.92483}              \\
      \textbf{\modelName} & \textbf{0.62488}$^\dagger$    & \textbf{0.82239}$^\dagger$    & \textbf{0.78168}$^\dagger$    & \textbf{0.76263}$^\dagger$ & \textbf{0.85398}$^\dagger$ & \textbf{0.92832}$^\dagger$ \\\midrule
      PLE                 & {\ul 0.60652}                 & 0.77282                       & 0.77157                       & {\ul 0.73986}              & 0.85640                    & 0.93391                    \\
      PLE + ii graph      & 0.59870                       & {\ul 0.80117}                 & {\ul 0.77185}                 & 0.73817                    & {\ul 0.85672}              & {\ul 0.93683}              \\
      \textbf{\modelName} & \textbf{0.62576}$^\dagger$    & \textbf{0.82238}$^\dagger$    & \textbf{0.78636}$^\dagger$    & \textbf{0.74725}$^\dagger$ & \textbf{0.85681}$^\dagger$ & \textbf{0.93897}$^\dagger$ \\\midrule
      Masknet             & 0.59360                       & 0.81263                       & 0.69159                       & \textbf{0.61034}$^\dagger$ & 0.82782                    & 0.86086                    \\
      Masknet + ii graph  & {\ul 0.63998}                 & \textbf{0.8166}$^\dagger$     & {\ul 0.72044}                 & 0.58009                    & {\ul 0.82889}              & {\ul 0.89715}              \\
      \textbf{\modelName}        & \textbf{0.65137}$^\dagger$    & {\ul 0.81631}                 & \textbf{0.72863}$^\dagger$    & {\ul 0.59534}              & \textbf{0.84161}$^\dagger$ & \textbf{0.90086}           \\ \bottomrule
    \end{tabular}}
\end{table}

\subsection{Online Performance Comparison} \label{on_comparison_exp}
To assess the effectiveness of {\modelName} in real-world industrial scenarios, online A/B Tests were conducted across the three recommendation scenarios in Alipay: Super 567, Consumer Channel, and Payment Result Page. Evaluation metrics differed based on the dataset. For Dataset A (Super 567) and Dataset C (Payment Result Page), both representing coupon issuance scenarios, we employed \#Click and \#Conv as evaluation metrics. \#Click denotes the number of coupons clicked by users, while \#Conv signifies the number of converted items. On the other hand, for Dataset B (Consumer Channel), which represents a selling goods scenario, we used \#Click and GMV as evaluation metrics. \#Click represents the number of goods clicked by users, while GMV (Gross Merchandise Volume) indicates the total monetary value spent by users on purchasing goods. Our objective was to increase both coupon conversion and goods GMV.

To conduct the A/B Tests, 10 percent of the actual online traffic was allocated, with the testing traffic candidates assigned randomly and evenly to two experimental groups. {\modelName} was compared against the online baseline approach, which represents the existing model version serving all online users. Over a period of one month, data on \#Click, \#Conv, and GMV were collected for the different experimental groups. The results of the online experiments are summarized in Table \ref{tab:online_experiment_result_all}. Due to commercial confidentiality, specific figures are withheld and represented with the symbol $``**"$. The percentage of relative improvement achieved by our method compared to the baseline is presented in the last row. The results demonstrate that our proposed {\modelName} approach achieved a 6.24\% increase and a 10.07\% increase in \#Conv for Dataset A and Dataset C, respectively. Additionally, a 6.45\% increase in GMV was observed for Dataset B. The results of the A/B Test demonstrate the significant improvements achieved by our method in real-world industrial recommendation scenarios.

\begin{table}[]
  \centering
  \setlength\tabcolsep{3pt} 
  \caption{The overall online performance comparison, where \#conv. is number of coupon conversion and GMV is Gross Merchandise Volume. Note that the improvements achieved by {\modelName} are statistically significant ($p$-value $\ll$ 0.05).}
  \begin{tabular}{l|cc|cc|cc}
    \toprule
                              & \multicolumn{2}{|c|}{Dataset A} & \multicolumn{2}{|c|}{Dataset B} & \multicolumn{2}{|c}{Dataset C}                                                                 \\
    \midrule\midrule
    Methods                   & \#Click                         & \#Conv.                         & \#Click                        & GMV                & \#Click            & \#Conv.             \\
    \midrule
    Baseline                  & 3* *7                           & 2* *0                           & 3* *0                          & 2* *1              & 7* *6              & 1* *1               \\
    Ours                      & 3* *7                           & 3* *6                           & 3* *9                          & 2* *2              & 7* *7              & 1* *2               \\
    \bfseries{\emph{Improv.}} & \bfseries{+2.67\%}              & \bfseries{+6.24\%}              & \bfseries{+6.18\%}             & \bfseries{+6.45\%} & \bfseries{+4.39\%} & \bfseries{+10.07\%} \\
    \bottomrule
  \end{tabular}
  \label{tab:online_experiment_result_all}
\end{table}

\subsection{Ablation Study}\label{ablation_exp}
In order to comprehensively evaluate the impact of the U2E2I recall module and the E-E-I model ranking module on {\modelName}, we conducted deeper ablation studies on Dataset A by selectively removing either the recall or ranking modules. The annotation \emph{w/o} indicates the absence of the U2E2I recall module or the E-E-I model ranking module, while \emph{w/} signifies the inclusion of these modules. The results are show in Table \ref{tab:online_ablation_experiment_result} and final row represents the improvements achieved by retaining each respective module compared to removing it. The experimental findings presented in Table \ref{tab:online_ablation_experiment_result} demonstrate that both U2E2I recall and E-E-I model ranking modules contribute to an increase in clicks and conversions, thus affirming the effectiveness of our U2E2I recall module and E-E-I model.

\begin{table}[htbp]
  \centering
  \setlength\tabcolsep{8pt} 
  \caption{The online ablation performance comparsion for Dataset C, where \emph{w/o} and \emph{w/} represent without and with, respectively.}
  \begin{tabular}{l|cc|cc}
    \toprule
    \multirow{2}{*}{Method}   & \multicolumn{2}{|c|}{U2E2I recall} & \multicolumn{2}{|c}{E-E-I model for ranking}                                        \\
                              & \#click                            & \#conv.                                      & \#click           & \#conv.          \\
    \midrule
    \emph{w/o}                & 7* *1                              & 1* *1                                        & 4* *7             & 8* *5            \\
    \bfseries {\emph{w/}}     & \bfseries {7* *7}                  & \bfseries{1* *2}                             & \bfseries {4* *5} & \bfseries{8* *2} \\
    \bfseries{\emph{Improv.}} & +3.33\%                            & +2.95\%                                      & +1.05\%           & +0.59\%          \\
    \bottomrule
  \end{tabular}
  \label{tab:online_ablation_experiment_result}
\end{table}

\subsection{Difference LLMs Comparison} \label{llm_exp}
In this subsection, we perform a comparative analysis of different large language models, namely ChatGPT, ChatGLM, and Claude. To assess their performance, we randomly selected 1,000 complementary entity pairs from the generated complementary graphs of these models. These entity pairs were manually evaluated and assigned scores based on their relevance. The scoring scale consists of five levels: 1-Completely unrelated, 2-Somewhat unrelated, 3-Uncertain, 4-Somewhat related, and 5-Completely related. The numbers of entity pairs falling into each of these five levels are reported in Table \ref{tab:compare_llms}. We then calculate the weighted average of these entity pairs using the following formula:

\begin{align*}
  \frac{1 \times \text{Completely Unrelated Num} + \ldots + 5 \times \text{Completely Related Num}}{1000}.
\end{align*}

This calculation yields the final manual judgment score, which is presented as the last row in Table \ref{tab:compare_llms}. Based on the manual judgment scores reported in Table \ref{tab:compare_llms}, it is evident that the complementary entity pairs recommended by Claude exhibit a higher level of correlation.

\begin{table}[htbp]
  \centering
  \setlength\tabcolsep{2pt} 
  \caption{Comparing the performance of complementary graph generated by different LLMs using five levels of manual annotation (randomly sampled 1000 entity pairs), a higher \emph{Mean Score} indicates that the model's predictions are closer to human judgments.}
  \label{tab:compare_llms}
  \begin{tabular}{@{}llll@{}}
    \toprule
    Model                    & ChatGLM 2 & ChatGPT 3.5 & Claude 2       \\ \midrule
    (1) Completely unrelated & 191       & 171         & 109            \\
    (2) Somewhat unrelated   & 40        & 26          & 36             \\
    (3) Uncertain            & 145       & 145         & 127            \\
    (4) Somewhat related     & 242       & 263         & 146            \\
    (5) Completely related   & 382       & 395         & 582            \\
    \midrule
    \bfseries{Mean Score}    & 3.584     & 3.685       & \textbf{4.056} \\ \bottomrule
  \end{tabular}
\end{table}

To provide a more comprehensive understanding, we also extract and present the instances of misjudgment made by ChatGPT and ChatGLM, where the models considered certain entity pairs as relevant, but they were manually determined as irrelevant. These instances are listed in Table \ref{tab:compare_llms_bad_cases}.

An analysis of the table reveals that ChatGPT associates ``Presbyopic Glasses" with ``Makeup Remover" based on the reasoning that Makeup Remover needs to be carefully applied by hand, and using a Presbyopic Glass after makeup removal can provide enhanced observation of the facial skin condition. ChatGLM, on the other hand, links ``Cake" with ``Pajamas" by suggesting that people may wear pajamas while eating cakes at night. We consider these explanations provided by the language models to be excessively imaginative, as they forcefully establish connections between these entity pairs.

\begin{table}[htbp]
  \centering
  \caption{The instances of problematic complementary entity pairs generated by the large language models (LLMs) from their respective complementary graphs.}
  \label{tab:compare_llms_bad_cases}
  \resizebox{\linewidth}{!}{
    \begin{tabular}{@{}llp{5cm}@{}}
      \toprule
      LLM Model   & Entity Pairs                     & Bad Reason                                        \\ \midrule
      ChatGPT 3.5 & \makecell[c]{Presbyopic Glasses,                                                     \\ Makeup Remover oil} &  \makecell[l]{Makeup Remover oil needs to be\\ carefully applied by hand, and using a \\Presbyopic Glass after makeup removal\\ can provide enhanced observation of\\ the facial skin condition.} \\ \midrule
      ChatGLM 2   & Cake, Pajamas                    & \makecell[l]{People may wear pajamas while eating \\ cakes at night.}
      \\ \bottomrule
    \end{tabular}}
\end{table}

\subsection{Case Study}
\begin{figure}
  \centering
  \includegraphics[width=0.48\textwidth]{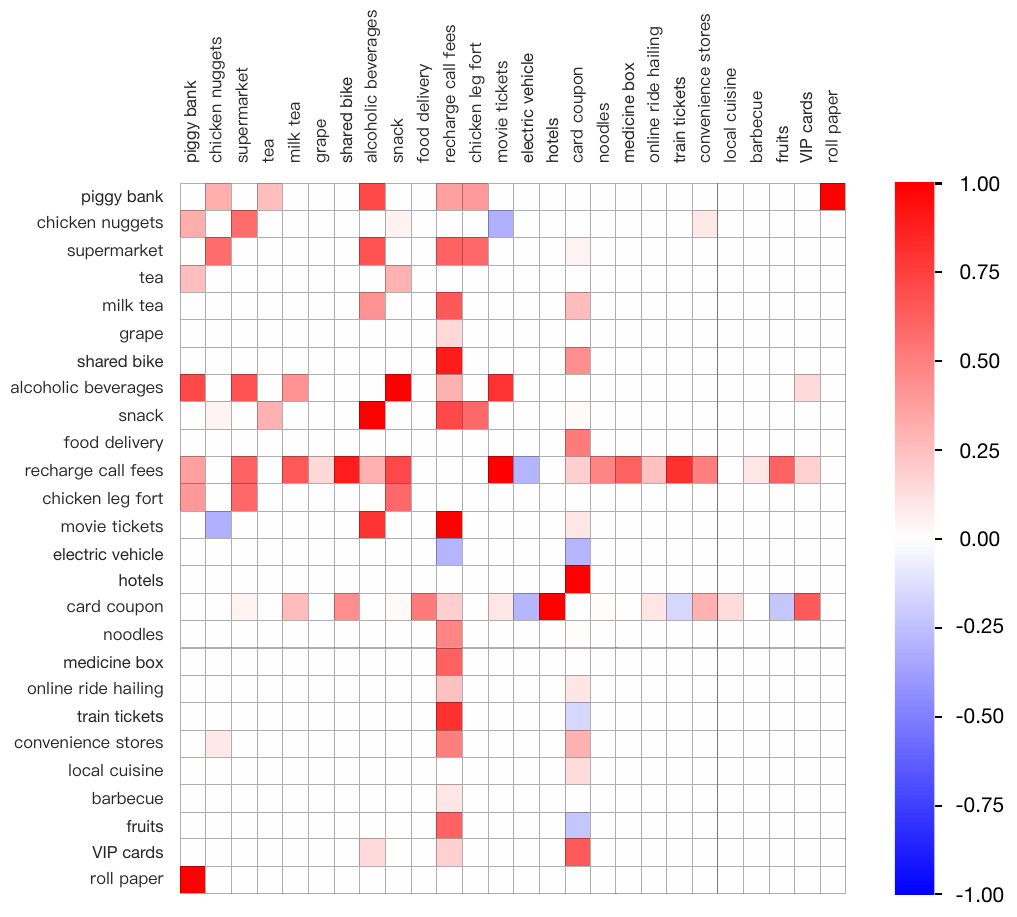}
  \caption{The relative improvement of conversion rate (CVR) for randomly sampled complementary pairs in {\modelName} compared to the baseline.}
  \label{fig:cvr_improve_compare_with_baseline}

\end{figure}

In this subsection, we present an additional case study focusing on the online experiment conducted on Dataset A. Specifically, we calculate and compare the Conversion Rate (CVR) of a sample set of complementary entity pairs recommended by {\modelName} and the baseline model. The comparison results are depicted in Figure \ref{fig:cvr_improve_compare_with_baseline}.
In the figure, blank squares indicate a non-associative relationship between the two entity words, while colored squares indicate the improvement in CVR of the experimental group compared to the baseline group. Red squares represent a higher CVR in the experimental group compared to the baseline group, while blue squares indicate a lower CVR in the experimental group compared to the baseline group.
As observed from Figure \ref{fig:cvr_improve_compare_with_baseline}, the complementary pairs recommended by the experimental group generally exhibit a higher CVR than those recommended by the baseline group.

%% file: related_work.tex










%% file: conclusion.tex
\section{Conclusion}

In this paper, we propose a novel LLM based Complementary Knowledge Enhanced Recommendation ({\modelName}) System.
It involves utilizing an entity extractor to extract unified concept terms from the information available for all items and user bills.
To construct a complementary graph, we initially generate entity pairs on their popularity and designed strategies. Next, we leverage a large language model to uncover existing complementary purchasing relationship between each entity pairs. 
Furthermore, we incorporate a new complementary recall module and train the E-E-I weight decision model to enhance the ranking model's knowledge and facilitate the recommendation of complementary items. 
Comprehensive experiments demonstrate the effectiveness of our proposed {\modelName} system.